\documentclass[a4paper]{article}
\usepackage{float}
\usepackage[fleqn]{amsmath}
\usepackage{indentfirst}     %Ê×ÐÐËõ½ø
\usepackage{epsfig}
\usepackage{CJK,graphicx}
\usepackage{multicol}
\usepackage{cite}
\usepackage{geometry}
\usepackage{color}
\usepackage{multirow}
\graphicspath{{My/}}
\begin {document}
\twocolumn[
\begin{@twocolumnfalse}
\title{Exponential distance distribution of connected neurons in simulations of two-dimensional in vitro neural network development}
\author{Zhi-Song $lv^{1}$, Chen-Ping $Zhu^{1,2}$*, Pei $Nie^{1}$, Jing $Zhao^{3}$,Hui-Jie $Yang^{2}$, Yan-Jun $Wang^{4}$, \\Chin-Kun $Hu^{5}$\\
1. Department of Physics in Science College, \\Nanjing University of Aeronautics and Astronautics, Nanjing, 210016, China\\
2. Research Center of Complex Systems Science, \\University of Shanghai for Science and Technology, Shanghai 200093, China\\
3. Department of Mathematics, College of Logistic Engineering of PLA, \\Chongqing 401331, China\\
4. College of Civil Aviation, Nanjing University of Aeronautics and Astronautics, \\Nanjing, 210016, China\\
5. Institute of Physics, Academia Sinica, Nankang, Taipei 11529}
%\date{}
\maketitle
\end{@twocolumnfalse}]
%\tableofcontents
\begin{abstract}
The distribution of the geometric distances of connected neurons is a practical factor underlying neural networks in the brain. It can affect the brain\'s dynamic properties at the ground level. Karbowski derived a power-law decay distribution that has not yet been verified by experiment. In this work, we check its validity using simulations with a phenomenological model. Based on the in vitro two-dimensional development of neural networks in culture vessels by Ito, we match the synapse number saturation time to obtain suitable parameters for the development process, then determine the distribution of distances between connected neurons under such conditions. Our simulations obtain a clear exponential distribution instead of a power-law one, which indicates that Karbowski's conclusion is invalid, at least for the case of in vitro neural network development in two-dimensional culture vessels.

\bf {Key words}: distance distribution, connected neurons, development, exponential, power-law, neural networks

PACS numbers:89.75.Fb
\end{abstract}
\section{ Introduction}
Complicated functions of the brain are largely attributed to the complex topological connections of a neural network (NN) that consists of about $10^{11}$ neurons and $10^{14}$ synaptic links. The geometric distance distribution of connected neurons in an NN is pertinent to its dynamic properties because it is related to real properties such as message transmission \cite{Chacron,Agnati}, energy cost \cite{Bullmore}, and response time \cite{Leveroni,Shahaf}. The measurement of such connections can be made at three levels: between brain regions \cite{Liang,Hagman}, between voxels, which are artificially divided blocks of brain tissue, and between neurons, which are tiny and hence difficult to discern. Experimental results have exhibited the small-world topology of structural connections at the first two levels.  At the microscopic level, nodes in the network are neuronal somas, while the links are synapses between a tip of the axon of one neuron and a terminal of a dendrite of another neuron. Large-scale in vivo measurements of the topological properties of an NN and the geometric distances between neurons are still challenges for researchers, although small-scale precise observation has been successfully performed
on C. elegance \cite{BJK} and mammalian visual cortexes \cite{Rotter}.
In contrast to in vivo experiments, significant achievement has been made on the in vitro development of NNs. For instance, Ito \cite{Ito1} found that synchronized bursts of an NN on a multi-electrode array in a cultured vessel required an initial plating density of at least 250 cells/$\ mm^2$ for Neuron Culture Medium and 500 cells/$\ mm^2$ for DMEM/serum. They also found that the final densities of surviving neurons at one month decreased greatly compared with the initial plating densities and stabilized in denser cultures. In another experiment, the densities of antibody-labeled synaptic terminals in both types of cultures increased gradually from 7 to 21-28 days in vitro (DIV). They did not increase further at 35 DIV and tended to become saturated \cite{Ito2}, which serves as a metric for numerical simulations to match.
From a theoretical aspect, Karbowski \cite{Karb} presented an optimal wiring principle and plateaus regarding the degree of separation for cortical neurons. He derived this result based on scaling theory from statistical physics using two neuron linking rules, minimal total axonal length and minimal total energy consumption during information transport, and let distance distribution $p( r )$ reside at the balance point between these two factors. By assuming that neurons in the cortex can be reached in just a few steps (i.e., the small-world property, which is now widely accepted), he obtained neural distance distribution $p( r ) \sim (r/\sigma)^z $ for $\frac{r}{\sigma}>>1$, i.e., the power-law decay for large distances with $z =  -2 \frac{1 - \alpha}{1-2 \beta}$ and $p(r) \rightarrow 1$ for $ r \rightarrow 0$. The quantity $\sigma$ denotes a microscopic length characterizing a neuron\'s size or the extent of local intra-cortical connections, where $\alpha < 1 $ and $\beta < \frac{1}{2}$, so that $z < 0$.
 Karbowski's power-law decay for distance distribution $p( r )$ between neurons has not been verified until now, as direct in vivo measurement in  human brains is not available. The purpose of our present work is therefore to check its validity in an indirect way. Based on a simple model, we simulate the in vitro development of rat NNs in culture vessels. Considering the death of isolated neurons and synapses formation process between connected neurons, we fit the model to both the saturation time of the number of synapses and neuronal density stabilization with the results of experiments done by Ito  \cite{Ito1,Ito2} by tuning the simulation parameters that describe various conditions. We then determine the distance distributions in the final networks that fit the development and saturation periods in Ito's experimental results well, so as to compare the Euclidean distance distributions obtained from the simulations with the power-law decay in Karabowski's prediction.
\section{Two-dimensional model for in vitro NN development}
To mimic in vitro NN development, we set up a simple model with the following rules. (1) Initial setting. We assume a cultured square vessel of size L with N neurons randomly scattered on its bottom plate. All coordinates of such nodes are recorded. Based on  Ito's experiments, the initial densities D = N/$L^2$ of the neurons are set to 500/$mm^2$ and 1000/$mm^2$, respectively, to ensure the formation of a final NN. The total culture time is 35 DIV. (2) Growth of axons and dendrites. We choose the soma radius of rat neurons to be $3 \times 10^3$ nm \cite{Miller}. Each axon of a neuron grows at a speed that is uniformly distributed in the range $(0, V_{max})$ in an arbitrary direction, where $V_{max}$ is an adaptable parameter to fit the experimental results and is assumed to be proportional to the vessel size. The axon grows $0$--$5$ branches after two weeks, and each branch is shorter than 1/5th the length of the axon, which corresponds with neuroscience textbook information. Each dendrite grows hierarchically with $2$--$4$ branches every three days after an initial growth of the first few days ($3$--$5$ DIV). We set its growth velocity to 130 nm/day, which is much smaller than $V_{max}$ , used for the axons \cite{Dotti}. In addition, we set its maximal length to 5000 nm, i.e., the terminal of a branch can generate $2$--$4$ new branches every three days after it has been generated. The center angle of the dendrites is parallel to that of the axon, and they span an angle of $240^o$, which is tunable. (3) Formation of synapses: When a tip of an axon is less than than 100 nm from a terminal of a dendrite, they form a synapse, because any synapse is actually a junction with a gap \cite{gap}. (4) The number of surviving isolated neurons $M(t)$ follows the rule \cite{Bullmore}:
$$M(t)  =  M(0)exp(-\lambda t),$$
where $M(0)$ is the number of remaining rat neurons isolated outside the NN and $\lambda$ is a decay constant in units of 1/DIV \cite{Bullmore}. In this phenomenological model, we have adjustable parameters $V_{max}$ \cite{Dotti}and $\lambda$ to fit the saturation time of the number of synapses in in vitro rat NN development \cite{Ito2}.
\section{Results of numerical simulations}
\subsection{Fitting synapse number saturation time in the development process}
 \begin{figure}[htbp]
 %Requires \usepackage{graphicx}
    \centering
    \includegraphics[width=3.25in]{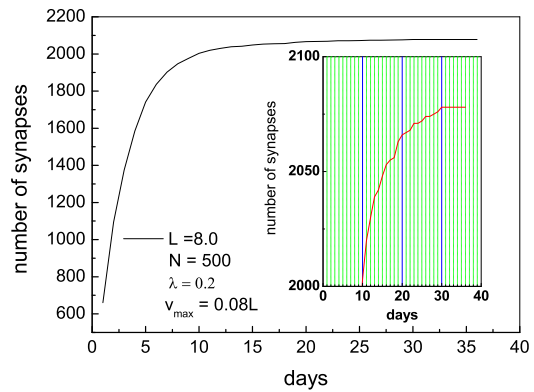}
\caption{Number of synapses $(S_a)$ increases with time first then gets saturated in growth development of neural network in vitro in cultured vessel. Simulation conditions: Vessel size $L = 8.0 mm$, neuronal density $D = 500$  cells/$mm^{2}$, decaying constant $\lambda = 0.2$/DIV, maximal growing velocity of axons $V_{max} = 0.08L$/DIV. Inset: enlarged curve of $S_a$ versus days(DIV) approaching saturation.}\label{fig1}
\end{figure}
\begin{figure}[H]
    \centering
    \includegraphics[width=3.5in]{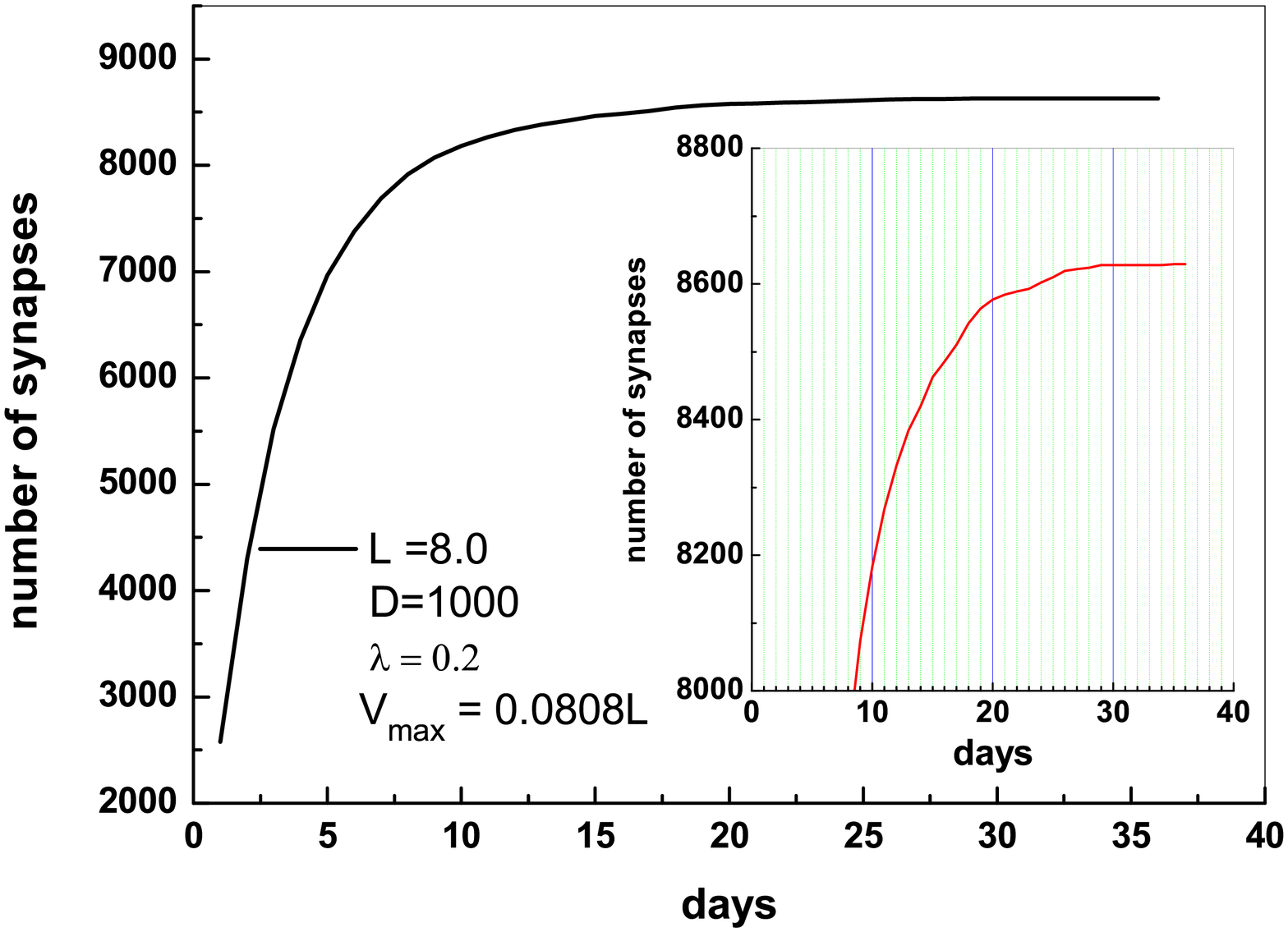}
\caption{Number of synapses $(S_a)$ increases with time first then gets saturated in growth development of neural network in vitro in cultured vessel. Simulation conditions: Vessel size $L = 8.0 mm$, neuronal density $D = 1000 $ cells/$mm^{2}$, decaying constant $\lambda = 0.2$ /DIV, maximal growing velocity of axons $V_{max} = 0.08L$/DIV. Inset: enlarged curve of $S_a$ versus days(DIV) approaching saturation.} \label{fig2}
\end{figure}
\begin{figure}[H]
    % Requires
    \centering
    \includegraphics[width=3.2in]{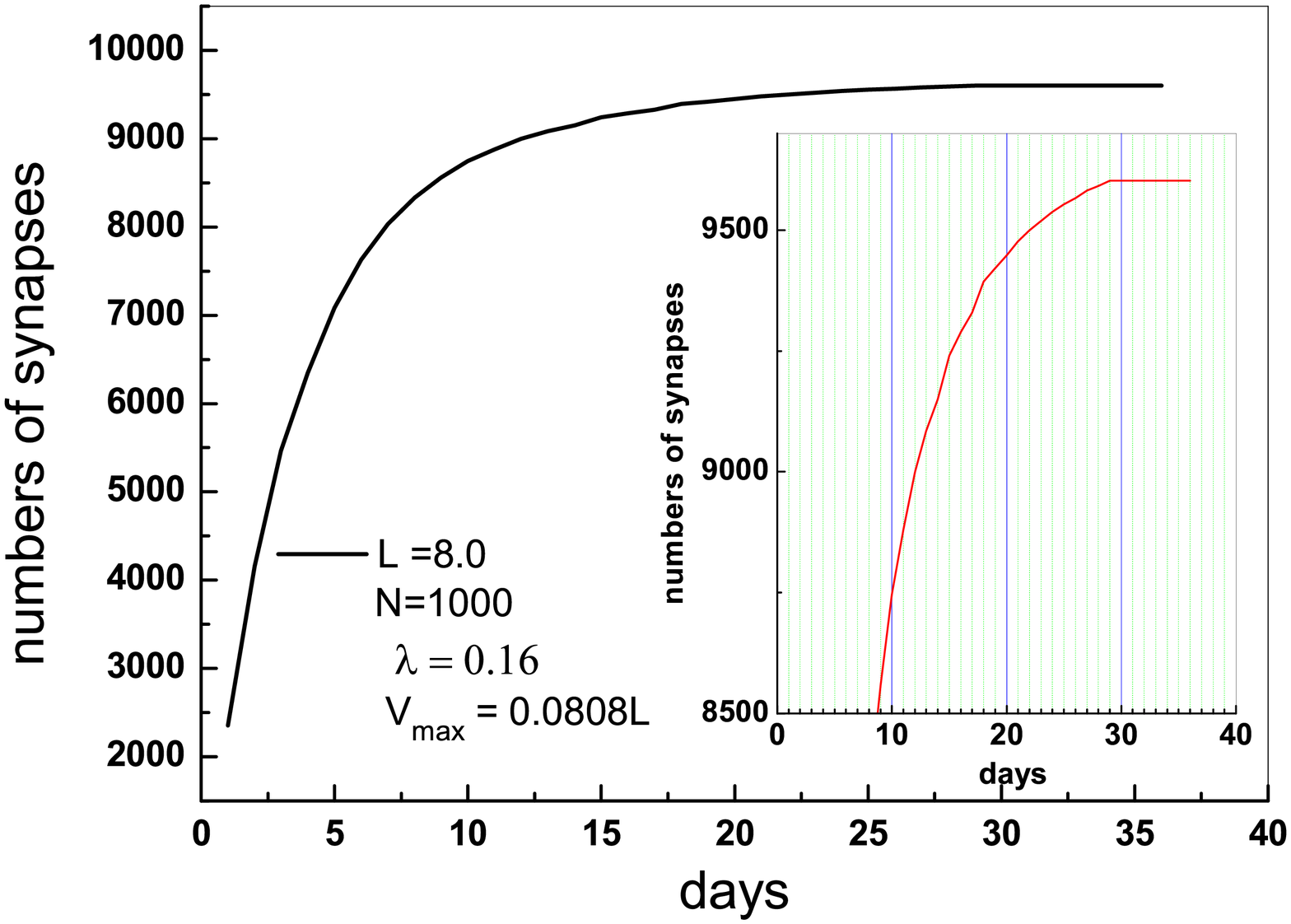}
\caption{Number of synapses $(S_a)$ increases with time first then gets saturated in growth development of neural network in vitro in cultured vessel. Simulation conditions: Vessel size $L = 8.0 mm$, neuronal density $D = 1000 $ cells/$mm^{2}$, decaying constant $\lambda = 0.16$ /DIV, maximal growing velocity of axons $V_{max} = 0.08L$/DIV. Inset: enlarged curve of $S_a$ versus days(DIV) approaching saturation. } \label{fig3}
\end{figure}
\begin{figure}[H]
    \centering
    \includegraphics[width=3.2in]{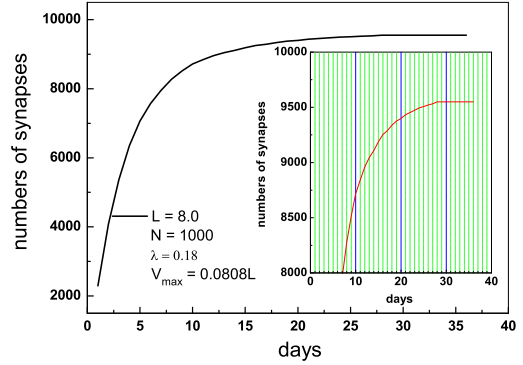}
    \centering
\caption{Number of synapses $(S_a)$ increases with time first then gets saturated in growth development of neural network in vitro in cultured vessel. Simulation conditions: Vessel size $L = 8.0 mm$, neuronal density $D = 1000$  cells/$mm^{2}$, decaying constant $\lambda = 0.18$ /DIV, maximal growing velocity of axons $V_{max} = 0.08L$/DIV. Inset: enlarged curve of $S_a$ versus days(DIV) approaching saturation.} \label{fig4}
\end{figure}
	
To ensure that the initially scattered neurons finally form a connected NN and the number of synapses saturates within the range of DIV reported in experimental results, we need to carefully adjust a parameter set that consists of decay coefficient $\lambda$ and maximal growth velocity $V_{max}$. The simulation results are not always good for arbitrary parameter values. For example, if we chose $\lambda = 0.1$ and $V_{max} = 0.04L$, $L = 4.0$ mm, with $D =500$/$mm^2$, no saturation could be found in the simulations.
Saturation time in the in vitro development experiment was fitted well by tuning parameters $\lambda$ and $V_{max}$ in the simulations. For a growth velocity in the range $(0, V_{max})$, where $V_{max} = 0.08L$, the number of saturated synapses is close to the empirical values using $\lambda = 0.2$ for culture vessels of sizes $L = 2.0, 4.0, $ and $8.0 $ mm. Typical results are shown in Figs.1 and 2. With an extended range of parameters, say $\lambda = 0.16$ or $0.18$/DIV, similar results were also obtained. Typical results are shown in Figs.3 and 4. One can see that the saturation times $S_a$ are $27, 29, 31$, and $29$ DIV for the parameters in Figs.1--4, respectively, which all agree with experimental data.
\subsection{Fitting the stabilization time of the number of surviving neurons in the development process}
\begin{figure}[H]
    \centering
    \includegraphics[width=3.2in]{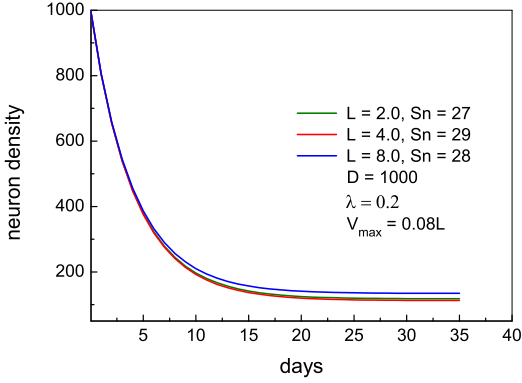}
    \centering
\caption{Number of survival neurons $(S_n)$ decreases with time first then gets stabilized in growth development of neural network in vitro in cultured vessel. Simulation conditions: Vessel sizes $L = 2.0, 4.0$, and $8.0 mm$, neuronal density $D = 1000$  cells/$mm^{2}$, decaying constant $\lambda = 0.2$/DIV, maximal growing velocity of axons $V_{max} = 0.08L$/DIV.}\label{fig5}
\end{figure}
\begin{figure}[H]
    \centering
    \includegraphics[width=3.2in]{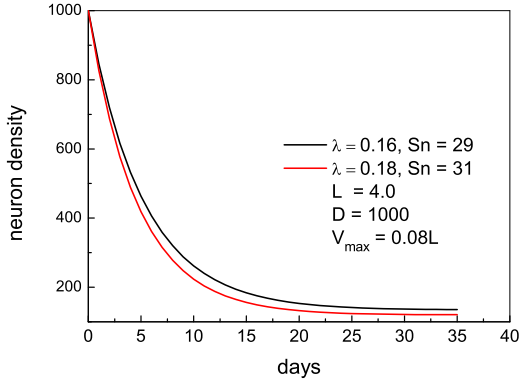}
    \centering
\caption{Number of survival neurons $(S_n)$ decreases with time first then gets stabilized in growth development of neural network in vitro in cultured vessel. Simulation conditions: Vessel size $L = 4.0 mm$, neuronal density $D = 1000$ cells/$mm^{2}$, decaying constant $\lambda = 0.16$ and $0.18$ /DIV, maximal growing velocity of axons $V_{max} = 0.08L$ /DIV.}\label{fig6}
\end{figure}
	
The number of surviving neurons decreases with time and arrives at a stable value when the number of synapses is saturated, which is obtained using the same modeling parameter pair in our simulations. Therefore, stabilization time $S_n$ for surviving neurons is estimated to be around $S_a$, which is supported by our simulations. Typical examples are shown in Figs.5 and 6 using the parameters $D = 1000$ /$mm^{2}$, and $V_{max} = 0.08L$ for $\lambda = 0.2$/DIV (Fig.5) and $\lambda = 0.16$ and $0.18$/DIV (Fig.6), respectively.
Stabilization time $S_n$ for the surviving neurons was found to be $ S_n = 27, 29,$ and $29$ DIV for $D = 1000$ cells/$mm^{2}$, $ \lambda = 0.2$ /DIV, and $V_{max} = 0.08L$ /DIV for $L = 2.0, 4.0$, and  $8.0$ mm, respectively. Further, $S_n = 29,$ and $31$ DIV for the same values of $D$ and $V_{max}$, $L = 4.0$ mm, and $\lambda = 0.16$ /DIV, and $0.18$ /DIV, respectively. One can see that the simulated stabilization time values are all close to those obtained in experiments, although stabilization time $S_n$ is not always the same as synaptic number saturation time $S_a$.
\subsection{Calculation of distance distribution between physically connected neurons in final NNs.}
\begin{figure}[H]
    \centering
    \includegraphics[width=3.2in]{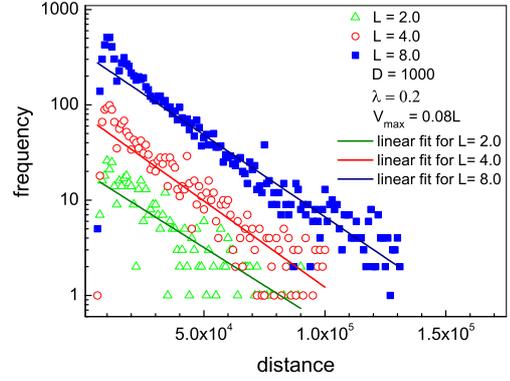}
    \centering
\caption{Exponential distance distribution $p( r )$ of connected neurons in final neural network in the conditions the same as Fig.2:  neuronal density $D = 1000$  cells/$mm^{2}$, decaying constant $\lambda = 0.2 $ /DIV, maximal growing velocity of axons $V_{max} = 0.08L$ but with different vessel sizes: $L = 2.0, 4.0$ and $8.0 mm$. }\label{fig7}
\end{figure}
\begin{figure}[H]
    \centering
    \includegraphics[width=3.2in]{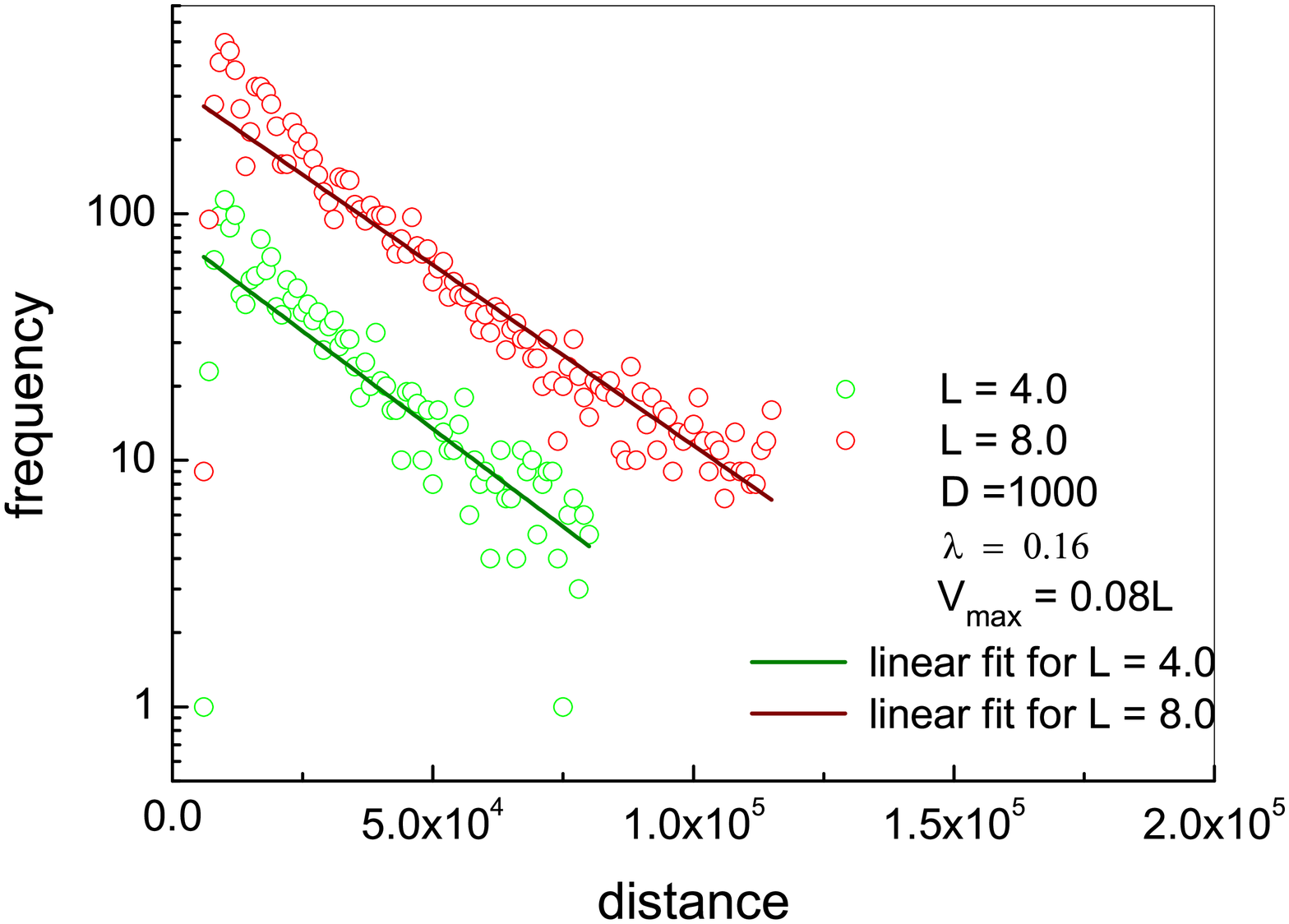}
    \centering
\caption{Exponential distance distribution $p( r )$ of connected neurons in final neural network in the conditions:  neuronal density $D = 1000$  cells/$mm^{2}$, decaying constant $\lambda = 0.16$ /DIV, maximal growing velocity of axons $V_{max} = 0.08L$ for different vessel sizes: $L = 4.0$ and $8.0 mm$.}\label{fig8}
\end{figure}
\begin{figure}[H]
    \centering
    \includegraphics[width=3.2in]{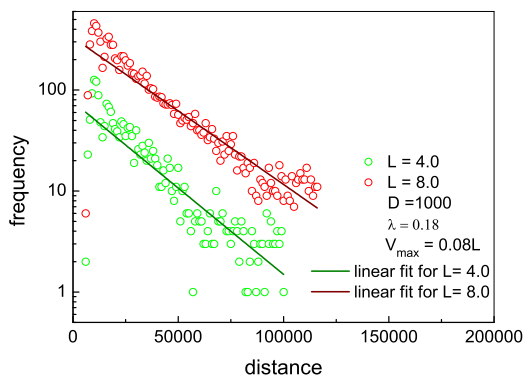}
    \centering
\caption{Exponential distance distribution $p( r )$ of connected neurons in final neural network in the conditions:  neuronal density $D = 1000 $ cells/$mm^{2}$, decaying constant $\lambda = 0.18$  /DIV, maximal growing velocity of axons $V_{max} = 0.08L$ for different vessel sizes: $L = 4.0$ and $8.0 mm$.}\label{fig9}
\end{figure}
Given the simulation results for which both synaptic saturation time $S_a$ and neuron number stabilization time $S_n$ fit well with experimental values, we can count the number of all possible distance values $r$ between all pairs of connected neurons in a NN after its saturation (stabilization) and calculate the distribution function of such distances. Such simulation results exhibit quite a definite exponential distance distribution: $p( r ) \sim e^{-\alpha r}$ for sufficiently large distance $r$, where $\alpha$ is a constant that relies on parameters of the culture vessels. Typical distributions of $p( r )$ are shown in Figs.7--9. With $\lambda = 0.2$ and $V_{max} = 0.08L $ for sizes $L = 2.0, 4.0$ and $8.0$ mm, we see an exponential decay tendency versus distance $r$, where $r >> \sigma$, for a fixed initial density of neurons $D = 1000$ cells/$mm^{2}$. Moreover, we see that larger numbers of neurons in the simulations lead to a clearer exponential distribution, which reveals the effect of statistical laws for finite systems. Similar results appear for the cases $\lambda = 0.16$ (Fig.8) and $\lambda = 0.18$ /DIV (Fig.9), each with $L = 4.0$ mm and $L = 8.0$ mm, respectively. All these results convince us that the distance distributions of NNs growing in vitro are exponential instead of following the power-law of Karbowski for the condition $r >> \sigma$.
In fact, such an exponential trend is obscured when growth velocity increases beyond a certain level $(V_{max}> 0.12 L)$, which limits the parameter range of the present model. Actually, the sizes of rat somas can change during development \cite{Miller}, but we use a simplified case with a steady size at the present stage. Moreover, when we checked our simulation range $(0, V_{max})$ with $V_{max} = 0.0808L$ for $L = 4.0$ mm and $L = 8.0$ mm, we found that velocities in these ranges correspond well with previously reported measured results \cite{Dotti}. In addition,  limiting simulation parameter $\lambda$ in the range $(0.16, 0.20)$ was inspired by experiments \cite{Bullmore}. These practical parameters ensure successful $S_a$ and $S_n$ fitting and extended results $p( r )$ of the proposed phenomenological model.
  All results simulated using parameters that mimic in vitro conditions in the culture vessels are listed in Table 1. The sizes of the culture vessels and the time needed to stabilize neuron numbers are not available from the reference. In this work, we increased sizes L while keeping the density D of neurons invariant. In this way, we repress fluctuations because of the finite size effect.
\section{Conclusions and Discussion}
To summarize, we obtained the necessary parameter values for both the range of axon growth velocity and decay rate of surviving neurons by fitting simulation results to the saturation (stabilization) time obtained from in vitro development in two-dimensional culture vessels. We then obtained an exponential distance decay law for connected neurons in NNs under such growth circumstances. The main point of this model is that surviving neurons grow their axons and dendrites ends to form synapses while neurons remaining isolated outside the network die out at a constant rate. Our simulation results do not support Karbowski's power-law decay function for physically connected neurons, although we cannot disprove it, as we do not know the range of the parameters in the brain, and this function was derived against a quasi-three dimensional background for the whole brain. However, because it is parameter-independent, which is quite different from what we have seen in experiments, the present simulation results indicate that it is invalid, at least for the development of NNs in two-dimensional culture vessels.
 Actually, the exponential decay distribution of distances between individual neurons in vivo has been revealed by both empirical investigation \cite{BJK} and theoretical models \cite{Kaiser,Ravasz,Kaiser2}. The present work simply contributes a phenomenological model for in vitro development in culture vessels. Indeed, nothing other than exponential decay can be obtained this way, as it contains a definite length scale (L). However, L-based axonal maximal growth velocity resides within the range of experimental results \cite{Dotti}. The exponential decay of the distance distribution actually implies a small-world network, in which the wiring cost, defined as the sum of the interneuronal distances, plays an important role in the actual neuronal network \cite{BJK}. It is characterized by dense local clustering (compared with that in scale-free networks) or the cliquishness of connections between neighboring nodes, yet a short path length between any (distant) pair of nodes is caused by the existence of a relatively few long-range connections. This is a model for the organization of anatomical and functional brain networks because a small-world topology can support both segregated/specialized and distributed/integrated information processing. Moreover, small-world networks are economical, tending to minimize wiring costs while supporting high dynamical complexity \cite{Bassett}. However, how the movable neurons in the brains of infants possibly rearrange \cite{Kaiser2} their positions to reach permanent sites, or if neurons in vitro really undergo such a rearrangement, remains a mystery for us to explore.
\section{Acknowledgement}
The anonymous referees are appreciated for their patience to review the manuscript
and for pertinent comments and suggestions for the revision.
The work is supported by Project No.11175086
of National Natural Science Foundation of China.
{}
\newpage
\begin{table}[H] \footnotesize %\scriptsize
\centering{}
\caption{A collection of simulated conditions and corresponding results with in vitro parameters used in experiments in ref. 10 and 11, where $S_a$ and $S_n$ represent saturated time of synapses numbers and stabilized time of neuron numbers, respectively, while Exp.$S_a$ and Exp.$S_n$ represent possible corresponding values available from experiments.}
\begin{center}
\begin{tabular}{cccccccccccc} %{p{1cm} p{2cm} p{0.7cm} p{1.cm} p{1.2cm} p{1.2cm} p{1.2cm} p{1.2cm} p{1.2cm} p{1.2cm} p{1.2cm} p{1.2cm} p{1.2cm}}
%\hline
\hline
& D(1/$mm^2$) & L(mm) & $N$ & $\lambda$$(1/DIV)$ & $V_{max}(1/DIV)$ & $S_a$(DIV) & Exp.$S_a$(DIV) & $S_n$(DIV) & Exp.$S_n$(DIV)\\
\hline																				
\multirow{7}{*}{}	&	 500 	&	 4.0 	&	8000	&	0.1	&	0.04L	&	{Not saturate}	&	$> 28$	& ¡¡Not stabilize& Not available	 \\
	&	 500 	&	 8.0 	&	32000	&	0.1	&	0.04L	&	{Not saturate}	&	$>28$		&	¡¡Not stabilize	&	NA	 \\
	&	 500 	&	 4.0 	&	8000	&	0.1	&	0.08L&	{Not saturate}	&	$>28$		&	¡¡Not stabilize	&	NA	 \\
	&	 500 	&	 8.0 	&	32000	&	0.1	&	0.08L	&	{	33	}	&	$>28$		&	Not simulated	& NA	 \\
	&	 500 	&	 4.0 	&	8000	&	0.2	&	0.08L	&	{	27	}	&	$>28$		&	Not simulated	&	NA \\
	&	 500 	&	 8.0 	&	32000	&	0.2	&	0.08L	&	{	30	}	&	$>28$		&	Not simulated	& NA \\
	&	 1000 	&	 2.0 	&	4000	&	0.2	&	0.08L	&	{	30	}	&	$>28$		&	27	&	NA \\
&	 1000 	&	 4.0 	&	16000	&	0.2	&	0.08L	&	{	27	}	&	$>28$		&	29	&	NA	 \\
&	 1000 	&	 8.0 	&	64000	&	0.2	&	0.08L	&	{	29	}	&	$>28$		&	28	&	NA	 \\
&	 1000 	&	 4.0 	&	16000	&	0.16	&	0.08L	&	{	31	}	&	$>28$		&	29	&	NA	 \\
&	 1000 	&	 4.0 	&	16000	&	0.18	&	0.08L	&	{	29	}	&	$>28$		&	31	&	NA	 \\
&	 1000 	&	 8.0 	&	64000	&	0.16	&	0.08L	&	{	29	}	&	$>28$		&Not simulated&	NA	 \\
&	 1000 	&	 8.0 	&	64000	&	0.18	&	0.08L	&	{	28	}	&	$>28$		&Not simulated&	NA	 \\
\hline
\end{tabular}
\end{center}
\end{table}
%\end{CJK}
\end{document}